\documentstyle[11pt]{article}
\newcommand{\blankline}{\vskip .3cm}
\newcommand{\f}{\begin{equation}}
\newcommand{\ff}{\end{equation}}

\begin{document}
\rightline{CGPG-96/93}
\blankline
\blankline
\blankline
\centerline{\LARGE Three dimensional strings as collective coordinates }
\blankline 
\centerline{\LARGE of four dimensional non-perturbative quantum gravity}
\blankline
\blankline
\rm
\centerline{Lee Smolin${}^*$}
\blankline
 \centerline{\it  Center for Gravitational Physics and Geometry}
\centerline{\it Department of Physics}
 \centerline {\it The Pennsylvania State University}
\centerline{\it University Park, PA, USA 16802}
\blankline
\blankline
\blankline
\centerline{September 7, 1996}
\blankline
\blankline
\blankline
\centerline{ABSTRACT}
A string theory in $3$ euclidean spacetime dimensions is found to 
describe the semiclassical behavior of a certain exact physical state 
of quantum general relativity in $4$ dimensions.   Both the worldsheet
and the three dimensional metric emerge as collective coordinates that 
describe a sector of the solution space of quantum general relativity.  
Additional collective coordinates exist which are interpreted as worldsheet 
degrees of freedom. The construction may be extended to the 
case in which the Kalb-Ramond field is included in the non-perturbative 
dynamics. It is possible that this mechanism is the inverse of the strong 
coupling limit by which some $D$ dimensional  string theories are 
conjectured  to give rise to $D+1$ dimensional field theories.
\blankline
${}^*$ smolin@phys.psu.edu
\eject

\section{Introduction}

In the last years quantum gravity has progressed dramatically
along two fronts.  At the perturbative level string 
theory\cite{strings} 
has been found to give a class of consistent
descriptions of the interaction of gravitons with matter.  
Recent results strongly suggest that this description may even
explain the thermodynamics of black holes\cite{strings-bh}.  There
are also many interesting results which suggest the existence of
a single non-perturbative theory from which all these 
perturbative descriptions arise\cite{stringduality,1011}.  However,  the 
proper setting
for this non-perturbative string theory  is still unknown,  
despite a number of interesting proposals.

At the same time, a non-perturbative description has grown
up for four dimensional quantum gravity based on the
Ashtekar-Sen variables\cite{abhay1} and their 
variants\cite{fernando-real,renata-real},
which comes from considering spaces of states constructed
from Wilson loops of the spacetime  connection
\cite{tedlee,pl,lp1,lp2,carlo-review,ls-review,aa-review,BGP,GP,g5}.  
Here there have been successes as well.
There are robust predictions about discreteness of geometrical
quantities at Planck 
scales\cite{ls-review,weave,volume1,AL1,L1,tt-length}.  
The spectra of the corresponding operators label a space of
diffeomorphism invariant states built from spin networks\cite{sn1}.
The dynamical
operators-the Hamiltonian constraint and Hamiltonian in fixed
gauges-are represented by gauge invariant and 
finite operators\cite{ham1,roumen-ham,ham2,qsdi,qsdii}.
Moreover, recently it has been shown that 
these results may be couched in a completely
rigorous formulation, giving non-perturbative quantum general
relativity a mathematical status on the level of constructive
quantum field theory\cite{AI,g5,AL1}
 
However,  there remain open problems, associated with the
form of the Hamiltonian constraint.  While the kinematical framework 
of the theory, at the
level of spatially diffeomorphism invariant states, seems
robust, there are many possible realizations of the Hamiltonian
constraint as a quantum operator.  These arise from different
regularization procedures.  Unfortunately, more than one of these
define theories that lack massless gravitons and the long ranged order 
required to have a limit in which general relativity is 
recovered\cite{instability}.
A new principle is clearly needed to pick out a form
of the Hamiltonian constraint that will naturally describe a
critical point at which a good classical limit may emerge.

It is natural then to wonder whether some marriage of perturbative
string theory and non-perturbative quantum gravity might
resolve their common problems and lead to a good
quantum theory of gravity.  Spin networks and the associated
observables (or some appropriate generalization of them) 
might provide a language for a non-perturbative
formulation of string theory, while string theory in turn might
suggest new forms for the non-perturbative 
dynamics that escape the problems
of quantum general relativity.    

There is, indeed, a strong
reason to suppose that non-perturbative quantum gravity
must involve string theory if it is to succeed. Suppose that a form of
non-perturbative quantum general relativity does exist which has
a good classical limit.  We should be able to use it to define a good,
Poincare invariant,
perturbative description of the interaction of gravitons on a
background Minkowski spacetime.  But after extensive search,
the only such consistent descriptions that are known are string
theories.  So, unless the search for perturbative quantum gravity
has missed something, the perturbation theory arising from the
semiclassical limit of {\it any} non-perturbative quantum theory
of gravity must be a string theory. 

But there are no consistent perturbative 
string theories whose low energy
limit describes only the graviton in four spacetime dimensions.
Therefore, it seems a consistent perturbative description
must have additional degrees of freedom.  They might arise in
three ways.  It might be necessary to add
supersymmetry and additional fields to the non-perturbative 
description for it to yield a good classical limit.  Alternatively, 
it may be necessary to apply the technology of
non-perturbative quantum gravity to $10$ or $11$ dimensional
theories. Or it might
be that additional degrees of freedom emerge directly from the
non-perturbative physics.

We may note that it is not completely crazy that a string theory
may emerge from a non-perturbative description of a theory 
based on Wilson loops.  There are suggestions that non-critical
strings may be associated with $QCD$.  Further, there is the
Klebanov-Susskind\cite{ks}  
construction which 
suggests that critical
string theory could emerge from a discrete theory based on
Wilson loops in a certain limit.  Non-perturbative quantum
gravity gives precisely such a discrete starting point based on
Wilson loops.

More generally, we may expect that to understand the emergence
of classical behavoir from the non-perturbative dynamics of spin
networks, we will have to discover collective coordinates that define
the large scale behavior of the physical states.  
To identify the collective
coordinates we must
study the systematics of generic spin network states that solve the
Hamiltonian constraint.  This is essentially what is done in this
paper.  The result is the identification of a set of collective coordinates
that describe the solution space of a certain sector of 
four dimensional non-perturbative
quantum general relativity.  A subset of these collective coordinates label 
imbeddings of
two dimensional surfaces in three dimensional metric manifolds.  
 
Using this result we can then construct a particular physical
state which has a semiclassical limit which describes the
propagation of a string in a three dimensional 
classical spacetime background.
The equations of motion of the string in this three dimensional
background turn out to follow from the condition
of stationary phase applied to this state.
Furthermore, there arise other collective coordinates which 
describe additional degrees of freedom that propagate on the surfaces.
The identification of these degrees of freedom, perhaps in terms of
conformal fields on the surfaces, has not yet been carried out.
Interestingly enough, the physics of these worldsheet degrees
of freedom depends on the form of the Hamiltonian constraint.

The results described here thus represent a first step in the
analysis of the collective coordinates of quantum general
relativity.  Even so, their form is interesting, as it is 
reminiscent of recent results that suggest that
string theories in $D$ dimensions may have 
strong coupling limits which are described by $D+1$ dimensional
field theories\cite{1011,34}.  It is then possible that the
mechanism described here implements the inverse process,
by which a $D$ dimensional string theory is recovered from
a weak coupling limit of a $D+1$ dimensional  
nonperturbative quantum field theory.

In the next section I describe how two dimensional surfaces emerge
as collective coordinates of quantum general relativity.  Section
3 introduces an exact solution which gives rise to string theory
in the semiclassical limit, which is the subject of sections 4 and 5.   
The Kalb-Ramond field is brought into the formalism in section 6.
In the
concluding section the question of whether these results can be
extended to supergravity in four and higher dimensions is discusssed.
The implications of these results are discussed in the conclusions after
which 
some technical issues associated with the choice of 
Hamiltonian constraint used here are discussed in the appendix.
A new proposal
for the form of the Hamiltonian constraint is described there as well.

\section{Two surfaces as collective degrees of freedom 
of quantum general relativity}

It is not difficult to see why two dimensional surfaces might naturally
emerge from the solutions to the constraints of quantum general
relativity.  The key points depend only on the basic features of 
non-perturbative quantum general relativity. 
The quantum
states of the theory have a basis which is diffeomorphism
equivalence classes of spin networks\cite{sn1}.  
A spin network is a graph, or more properly
the diffeomorphism class of a graph\footnote{Unless otherwise 
specified all references to spin networks in this paper are to  
diffeomorphism classes.  These are sometimes denoted
$\{ \Gamma \}$.},  whose edges are labeled
by spins such that the laws of addition of angular
momentum are satisfied at the vertices\cite{sn-roger}. 
The dynamics
is enforced by the Hamiltonian constraint, which is known
to act only at the nodes of the spin 
networks\cite{tedlee,lp1,ham1,roumen-ham,renata2,GP2,qsdi,qsdii}.  
Given a
certain definition of that operator, its action is
intrinsically planar.  This is because the action of the
constraint on a spin network is to extend  the network by
creating new {\it trivalent vertices} in a particulr way. This planar action
of the constraint means that there are classes of solutions constructed
from infinite superpositions of spin networks each of which 
span a given two 
dimensional surface.  

I will assume that the action of the Hamiltonian
constraint is according to a set of simple rules.
These may be derived from one particular
regularization\cite{ham1,roumen-ham,ham2} 
but  this is not unique;  there are
other regularizations that will also lead to them.   It is also possible
to modify the rules I give slightly while preserving the most
interesting results.  So as not to
bore non-experts, these and other technical issues associated with
the different regularizations of the hamiltonian constraint
are treated in the appendix.

\begin{itemize}

\item{\bf R1}  The hamiltonian constraint ${\cal C}(N)$ acts
at each node of a spin network.
The action at each node is a sum of terms,  
one for  every pair of its non-collinear edges.  
In each term two new nodes are added, each
on one of the two edges.  These nodes are placed adjacent
to the node at which the constraint acts.  The two new nodes
are then joined by a new edge with color $1$. (We use as is
traditional with spin networks a labeling in terms of twice
the spin, so all labels are integer multiples of $\hbar/2$.). Thus
the two nodes created by the action of the constraint are
trivalent.  
The diffeomorphism class of the 
edge is chosen so that the triangle it forms with the
two existing edges
links no other edge of the graph.
For each such term a linear sum of  
four new spin networks is then created, in each of which
the colors of the edges joining the old node to the new nodes
are raised or lowered by $1$.  The action of the
Hamiltonian constraint is to produce these four new
spin networks for every pair of non-collinear edges at every node
of the graph, multiplied by coefficients,
$A^{\pm \pm^\prime }(i,j,k)$, where $i$ and $j$ are
the spins of the two edges acted upon, and $k$ is the
third spin at the node\footnote{If the node has more than
three incident edges we decompose it into a product of trivalent
nodes as described in \cite{sn1} such that there is an internal edge
that connects the two external edges labeled $i$ and $j$ to the 
rest of the node. $k$ is then the color of this internal edge.}.  
The formula for the coefficient
is given by 
eq. (25) of \cite{roumen-ham}, but its exact form will not
be relevant here.   The two
new vertices that are created are trivalent, and consist
of the line of the new edge joining one of the old edges.
 
\item{\bf R2}  The action of the Hamiltonian constraint on each node is
multiplied by an arbitrary  factor $N$  at each distinct node.  This
means that the solutions must be found independently for
the action at each node.     
However, as we are giving the action on diffeomorphism
invariant states we must give a rule that tells how distinct
nodes are to be identified.  This must respect diffeomorphism
invariance, but reproduce the results of the standard action.
The following rule, which makes use of the graph recognition
problem\cite{instability}, works\footnote{One might worry that there 
are many examples in which
a given graph $\Gamma$ may be identified in more than
one way with a subgraph of $\Gamma^\prime$, or
that the subgraphs may have symmetries that prevent the
unique identification of the node.  However, the combinatorics
of the graph recognition problem tells us that the proportion
of such cases goes to zero rapidly as the graphs become large.
As we are interested in the classical limit, and hence 
large complex graphs, this is sufficient.}.  

We multiply the action  of the
operator on each node ${v}$ by numbers
$N({v})$, which are assumed to be assigned independently
to all nodes of all networks, subject to the following
restriction.  When it is the case that a network
$\Gamma$ may be identified as a sub-network of $\Gamma^\prime$,
such that a given vertex $v$ of $\Gamma$ is identified uniquely
with a vertex $v^\prime$ of $\Gamma^\prime$
then $N(v)$ in the action on
$\Gamma$ must be taken equal to $N(v^\prime)$ of
$\Gamma^\prime$.

\item{\bf R3} The Hamiltonian constraint acts thus at every
pair of non-collinear tangents of every node,  bivalent
trivalent or higher, independent of whether the tangent
vectors of the incident edges span the tangent space 
at the node.

\end{itemize}

We will shortly find reason to modify this last rule slightly, but it
is useful to begin the discussion with this form.

\subsection{How collective coordinates arise from the dynamics of
spin networks}

In order to extract the collective coordinates of a theory we must
investigate the systematics of its solutions.  The generic
solution to the Hamiltonian constraint of quantum general relativity
is constructed from an infinite superposition of spin network
states.  To understand the properties of the solution, we may 
investigate the trajectories generated by the Hamiltonian constraint
in the space of spin network states, $\cal S$.  To do this let us imagine that
we begin with some initial network $|\Gamma >$ and act on it with
an infinite number of iterations of the Hamiltonian constraint, with
arbitrary values of the lapses $N$, constrainted by {\bf R2}.  The
result is a subspace ${\cal F}_\Gamma$.  We would like to investigate
some general properties of these subspaces, as solutions are going
to be constructed from superpositions of states inside each of them.  

The most important observation is that with the rules given above
(and in most definitions of the Hamiltonian constraint in the
literature\cite{tedlee,lp1,ham1,roumen-ham,ham2,qsdi,qsdii})
all trajectories converge on a particular subspace, ${\cal S}^3$, which is
spanned by spin networks with only trivalent nodes.  
This is because the Hamiltonian constraint in all
these forms creates only trivalent vertices.  Thus, 
${\cal S}^3$ is an attractor for the orbits of the Hamiltonian
constraint.  

For this reason all considerations of this paper will be restricted
to either ${\cal S}^3$ or ${\cal S}^{2,3}$, the subspace consisting  of
networks with only trivalent and bivalent nodes\footnote{Bivalent
nodes are often called kinks}.  
One may regard this as a simplifying
assumption, but it may also be that the microscopic theory
restricted to the trivalent sector may be sufficient to define a good
quantum theory of gravity.  Some reasons to expect this are 
described in the appendix\footnote{To the obvious objection that this sector
is degenerate because all states have zero volume one may
reply that once the theory is quantum deformed\cite{sethlee}, this is
no longer the case\cite{rsl}.  This is discussed further below.}.  

A second key observation is that restricted to the 
trivalent sector the action of the  Hamiltonian 
constraint defined by the above rules 
will  produce graphs which are topologically planar.   
This means that for every two dimensional surface
$S$ there is a collection of orbits of the hamitonian constraint 
 such that each initial network $\Gamma$ and, every 
every spin network in its orbit,  is a skeletonization
of  $S$.  This is how two dimensional surfaces arise as collective
coordinates for solutions of the Hamiltonian constraint.

Once this is seen, one can observe also that the general orbit may
be described in terms of a two dimensional surface imbedded in a link
consisting of some number of non-intersecting loops (as these do not
evolve\cite{tedlee}).  This will enable us to define a semiclassical limit
in which the surface is embedded in a three dimensional metric, defined
by the link.  

We now proceed to describe how these results are obtained.

\subsection{States associated with open surfaces}

We begin with the simplest example of a graph that
spans an open surface.  Let us take for the initial spin network
$\Gamma$ an unknotted
triangle whose three edges each have spin $j$.

We will call the associated diffeomorphism invariant
state $|\Delta , j>$.  Let us
describe the subspace ${\cal F}_{\Delta , j}$ generated
by repeated application of the Hamiltonian constraint.  The first
time it acts, the Hamiltonian constraint will produce $12$ planar graphs,
each of which has one corner of the triangle bisected.  
($12=3$ corners times four graphs made at each corner.)  
If we act again with the Hamiltonian constraint we now 
produce $28=7 \times 4$ new graphs, as there are now seven
places where the Hamiltonian constraint will act.  If we
continue to iterate the action we will construct an infinite
family of planar spin networks, whose boundaries are
each the original triangle.  These make a subspace of
the diffeomorphism invariant state space associated to the
original triangle, which we call ${\cal F}_{\Delta , j}$.
This is a graded space, where each subspace 
${\cal F}_{\Delta , j}^n$ consists of graphs produced
by $n$ actions of the Hamiltonian constraint.
We will label the elements in each 
graded subspace by an arbitrary index
$\alpha$, so the states in ${\cal F}_{\Delta , j}$
will be called $|\Delta , j ; n, \alpha >$.

It is clear that we should find a set of solutions to the
constraints inside of ${\cal F}_{\Delta , j}$.    There
are a set of matrices ${\cal M}^n_{\alpha \beta}$ which
are defined by 
\f
{\cal C}(N) |\Delta , j ; n, \alpha > = 
\sum_{\hat{n} \in |\Delta , j ; n, \alpha >} \sum_{\beta}
N(\hat{n} ) {\cal M}^n_{\alpha \beta} |\Delta , j ; n+1, \beta >
\ff
where $\beta$ includes also the fact that we sum over the
raisings and lowerings of the spins of the edges and
the matrix elements contain the coefficients
$A^{\pm \pm^\prime }(j,k)$.

To find non-trivial solutions we will have
to  use a Hermitian ordered
operator corresponding to the constraint.  This means we
are interested in the constraint
\f
{\cal H}(N)= {\cal C}(N) + {\cal C}^\dagger (N)
\ff
where the hermitian conjugate is taken in the inner product
defined for the Euclidean theory in \cite{ls-review,g5,dpr}.
In this inner product, the distinct diffeomorphism classes
of spin networks comprise an orthogonal basis.  
 
To define the Hermitian conjugate 
we also have to know what we mean by 
$ N(n ) $, where $n$ is the position of a node.
We will assume that this is defined on diffeomorphism
invariant states according to
rule {\bf R2}.   

With this provision, the adjoint of the Hamiltonian
constraint acts in the following way.  It searches the graph
for nodes which have adjacent to them two vertices connected
by an edge with spin $1$, of the type created by the 
Hamiltonian constraint.  This means that these vertices
have two collinear edges which differ by $\pm$ one unit
of spin.  The operator than removes the edge, sets the
spins on the two edges to agree with those outwards of the
two vertices we have just removed and multiples by 
an associated  coefficient  $B^{\pm \pm^\prime }(j,k)$.
It then multiples by $N(\hat{n})$ at that vertex.

In equations,  there are matrices 
${\cal N}^n_{\alpha \beta}$ such that
\f
{\cal C}^\dagger (N) |\Delta , j ; n, \alpha > = 
\sum_{\hat{n} \in |\Delta , j ; n, \alpha >} \sum_\beta
N(\hat{n} ) {\cal N}^n_{\alpha \beta} |\Delta , j ; n-1, \beta >
\label{adjointaction}
\ff
We may note that acting on any state in ${\cal F}_{\Delta,j}$ the
adjoint ${\cal C}^\dagger (N)$ produces a state which is
also in ${\cal F}_{\Delta,j}$.  Given the standard inner product
in which the distinct spin networks are orthogonal this is true
for any ${\cal F}_\Gamma$ as long as 
\f
{\cal C}^\dagger (N) |\Gamma >=0
\label{rootcondition}
\ff

We may then construct solutions inside of ${\cal F}_{\Delta , j}$
by expanding them as
\f
|C > = \sum_n \sum_\alpha C(n,\alpha ) |\Delta , j ; n, \alpha >
\label{statesum}
\ff
We look for solutions such that
\f
{\cal H}(N)|C> = \left ( {\cal C}(N) + {\cal C}^\dagger (N) \right ) |C> =0
\label{hamconstraint}
\ff
To find these we note that the matrices ${\cal M}^n_{\alpha \beta} $
and ${\cal N}^n_{\alpha \beta}$ may be described in more detail
in terms of the nodes.  Each nonzero element of  
${\cal M}^n_{\alpha \beta} $   comes from
an $\alpha$ and a $\beta$ where  $|\Delta , j ; n+1, \beta >$
is gotten from $|\Delta , j ; n, \alpha >$ by dressing some node
$\hat{n} \in |\Delta , j ; n, \alpha >$.    If we label arbitrarily
the dressings by an index $d$ (which include the
raising and lowerings of spins) then for each nonzero
element there is a node $\hat{n} \in |\Delta , j ; n, \alpha >$
and a dressing $d$ such that $\beta = f(\alpha ,\hat{n} , d)$.
Thus, we can write
\f
{\cal C}(N) |\Delta , j ; n, \alpha > = 
\sum_{\hat{n} \in |\Delta , j ; n, \alpha >} \sum_d
N(\hat{n} ) {\cal M}^n_{\alpha \beta (\alpha , \hat{n},d) } 
|\Delta , j ; n+1, \beta (\alpha , \hat{n},d) >
\ff
Similarly, in \ref{adjointaction} 
the sum over $\beta$ is nonvanishing
only over those $\beta=g(\alpha , \hat{n} )$ 
gotten from $\alpha$ by removing edges and pairs of
nodes according to the procedure just described.  Note
that there is no dressing parameter as there is always
a unique way to undress a node, because in a trivalent
graph there is always at most
one unique edge whose two ends are adjacent to a node.
Thus, 
\f
{\cal C}^\dagger (N) |\Delta , j ; n, \alpha > = 
\sum_{\hat{n} \in |\Delta , j ; n, \alpha >}
N(\hat{n} ) {\cal N}^n_{\alpha \beta (\alpha, \hat{n} )} 
|\Delta , j ; n-1, \beta (\alpha , \hat{n}>
\ff
Hence, we have then to solve the
set of equations,
\begin{eqnarray}
0&=&\sum_n \sum_\beta  |\Delta , j ; n, \beta >
\left ( \sum_\alpha \sum_{\hat{n} \in |\Delta , j ; n, \alpha >}
\sum_d C(n-1 ,\alpha ) N(\hat{n} ) 
{\cal M}^{n-1}_{\alpha \beta } \delta_{\beta f(\alpha, \hat{n}, d )}
\right.
\nonumber \\
&&+ 
\left. \sum_\alpha \sum_{\hat{n} \in |\Delta , j ; n, \alpha >}
C(n+1 ,\alpha ) N(\hat{n} ) 
{\cal N}^{n+1}_{\alpha \beta } \delta_{\beta g(\alpha, \hat{n} )}
\right )
\end{eqnarray}
This gives us an independent equation to solve at every
node of $\beta$.  To see this note that in the first sum,
$|\Delta , j ; n-1, \alpha >$ is necessarily a subgraph
of $|\Delta , j ; n, \beta >$ (to say this we allow graphs to
have a spin zero edge).  The sum over nodes can then be
extended to a sum over nodes of $\beta$. We need only
add the notion that the function $f(\alpha, \hat{n}, d )$
returns a trivial graph in the case that $\hat{n}$ is not
a node of $\alpha$.  In the second sum the situation
is reversed, $|\Delta , j  ; n-1, \beta >$ is  a subgraph
of $|\Delta , j  ; n, \alpha >$.  But a node that is in 
$\alpha$ but not in $\beta$ would never contribute to the
sum, because the node refers to one that is undressed,
which means it remains.  (The nodes removed are two that are 
adjacent to it. )  
Hence we can extract the sum over nodes, so we have
\begin{eqnarray}
0&=&\sum_n \sum_\beta  |\Delta ,j ; n, \beta >
\sum_{\hat{n} \in |\Delta , j ; n, \beta >} N(\hat{n} )
\left ( \sum_\alpha  
\sum_d c(n-1 ,\alpha )  
{\cal M}^{n-1}_{\alpha \beta } \delta_{\beta f(\alpha, \hat{n}, d )}
\right. \nonumber \\
&&+ \left.
\sum_\alpha  
c(n+1 ,\alpha )  
{\cal N}^{n+1}_{\alpha \beta } \delta_{\beta g(\alpha, \hat{n} )}
\right )
\end{eqnarray}
Hence, for every $\beta$ and 
$\hat{n} \in |\Delta , j ; n, \beta >$ we must solve,
\f
0= \left ( \sum_\alpha  
\sum_d C(n-1 ,\alpha )  
{\cal M}^{n-1}_{\alpha \beta } \delta_{\beta f(\alpha, \hat{n}, d )}
+ \sum_\alpha  
C(n+1 ,\alpha )  
{\cal N}^{n+1}_{\alpha \beta } \delta_{\beta g(\alpha, \hat{n} )}
\right )
\ff
This gives us solutions, parameterized by the $C(n,\alpha )$.
We see that taking the Hermitian part is necessary to get solutions,
what is happening is that the amplitude of a graph created
by the action of ${\cal C}$ by adding to its subgraphs must be
balanced by the amplitude for ${\cal C}^\dagger$ to create the
same graph by removing an edge from a graph it is a subgraph
of.  Nothing is known presently about the 
space of solutions or its parameterization.
I will therefor just parameterize them by a symbol $Z$, so the
solutions to \ref{hamconstraint} given by 
\ref{statesum} are called $|\Delta , j ;  Z >$.
Thus, we have a space of states associated to a planar
surface, bounded by a triangle, each of which presumably
consists of an infinite sum over spin network states that
share the same triangle as its boundary.  (Note however that
in every state in this sum the spins on some part of the
boundary will differ from the original assignment.)
We shall call this the space of physical states
${\cal P}_{\Delta , j } \in  {\cal F}_{\Delta , j}$.

One property of the solutions is clear by inspection.  The linear
equations that must be solved separate into three sets,
each associated with the dressings of one of the nodes of
the triangle.  This is an aspect of the problem of confined 
correlations,
which is discussed at length in \cite{instability}.  
Thus, each solution may be visualized in the following way.  
Consider three disks, labled
by $\alpha=1,2,3$, each of which is joined to the two
others by an edge labeled by $j$.  Each of the disks stands
for an infinite superposition of spin networks which span the
disk topologically, each with two external edges with spin $j$.
Each disk may be labeled with a parameter $Z_\alpha$ that
parameterizes the solution in the neighborhood of the 
$\alpha$'th node of the initial triangle.  
These $Z_\alpha$ may be considered to be additional collective
coordinates that describe the solutions.   

It is clear that this procedure can be immediately generalized
to construct the spaces ${\cal F}_\Gamma$ that are generated
by acting on any initial trivalent spin network $\Gamma$,
as long as \ref{rootcondition} is satisfied.  
We now proceed to discuss the more interesting cases that
are associated with closed surfaces.

\subsection{States associated to closed surfaces}

It is clear that a similar construction also lets us associate
spaces of states with closed surfaces. For example, we may
construct planar states with the topology of $S^2$.  To make
the simplest example, consider first a simple tetrahedron,
${\cal T}$ whose edges are dressed with spins
$j_i$, $i=1,...,6$.  We will call the associated diffeomorphism
invariant spin network 
state $|{\cal T}, j_i >$.  Associated to it we have the  infinite
dimensional space  of states that we may get by acting 
an arbitrary number of times with the Hamiltonian
constraint, which we  call
${\cal F}_{{\cal T}, j_i}$.  An element of which will be
labeled $ |{\cal T}, j_i ,n ,\alpha>$ as before.   
These states have properties similar to the ones
that we described that dress triangles.  In each state
each triangle is dressed as before, but along each edge
one finds various patterns of edges that participate in the
dressings of one or the other of the edges it bounds.   

In just the same way we can solve the Hamiltonian constraints
inside ${\cal F}_{{\cal T}, j_i}$, giving us a space  of
physical states $ |{\cal T}, j_i ,Z >$ which each consists of 
linear combinations of states that are topologically $S^2$.
In this same way, given any initial trivalent spin network $\Gamma$
that satisfies \ref{rootcondition}
we can construct a space of physical states 
${\cal P}_{\Gamma} \subset {\cal F}_\Gamma$.   
 
As in the case of the triangle, the solutions are independent
in the neighborhoods of each node of $\Gamma$, because
the hamiltonian constraint cannot is block diagonal
and does not mix labelings of the edges that it creates outside
of these neighborhoods.  Thus the solutions can be pictured as 
a collection of 
disks, each of which has two or three external edges, tied together
with the topology of the initial network $\Gamma$.  Essentially what
the Hamiltonian constraint has done is to grow each node of 
$\Gamma$ into one of these disks.  The disks are connected
by edges with the same spins as in $\Gamma$ and they 
are labeled by
parameters $Z_\alpha$ which are collective coordinates that parameterize
the solutions.

We can then extend the disks, joining them in each $n$-gon of
$\Gamma$ untill they form a continuous surface, $S$.  
This surface will in general be self-intersecting, and it may
be open or closed. 

Inversely, we may begin with a surface $S$ and consider
the state spaces ${\cal F}_{S} = \oplus {\cal F}_\Gamma$
such that $\Gamma$ is a skeletonization of
$S$ satisfying \ref{rootcondition}, and the
corresponding spaces of physical states,
${\cal P}_{S} = \oplus {\cal P}_\Gamma$.  
Associated to each of these there must
as well be an algebra of physical observables ${\cal A}_S$,
which is a subalgebra of observables of quantum gravity and
which serve to distinguish all the states in ${\cal P}_S$.
 This algebra describes a quantum field theory associated
with the surface $S$, whose degrees of freedom include the
collective coordinates $Z$.  

There is also an additional degree of
freedom, associated with the choice of the initial $\Gamma$ that
spans $S$.  
Thus, the collective coordinates
that describe the solutions we have constructed
are the triples  $(S,\Gamma , Z)$.   

Before turning to a study of the classical limit, it is 
appropriate to make several comments.

First, with the rules given, 
the collective coordinates $Z_\alpha$ are
non-propagating as they are describe
degrees of freedom associated to each of the disks gotten by
solving the contraints in a neighborhood of each vertex of the
initial network $\Gamma$.  This means they cannot be described by 
a massless field theory on $S$.  It would be much preferable
to eliminate this problem so that the collective coordinates $Z$
described degrees of freedom which could propagate over the
whole surface, as this means it might be possible to represent
them in terms of a conformal field theory.  A set of rules that 
accomplishes this is described in the appendix.  

Second, it is easy to modify the theory so that there are
no solutions associated to 
open surfaces.  We need only modify assumption {\bf R3} to read:

\begin{itemize}

\item{\bf R3'} The Hamiltonian constraint acts thus at every
pair of non-collinear tangents of every node, 
trivalent or higher, independent of whether the tangent
vectors of the incident edges span the tangent space at the node.
{\it It does not act at bivalent vertices.}

\end{itemize}

How this modification may be accomplished is discussed
in the appendix.
I will assume from now on that a regularization of
the Hamiltonian constraint satisfying {\bf R3'} is made,
so the surfaces are all closed.  As a result, from now 
on we restrict attention to the space ${\cal S}^3$ whose
states have only trivalent vertices joining smooth edges.  

Third, if we adopt the rule {\bf R3'} 
it is straightforward to 
describe the general solution in the trivalent sector
${\cal S}^3$.
The reason is that any trivalent spin network 
$\rho$ can be
decomposed into a set of connected trivalent graphs 
which we will call $\Gamma$ and a link, which consists of
a set of non-intersecting loops, which we will call 
$\gamma$.   
In what follows, these symbols will stand for the pairs of the
graphs and the spins that label their edges.
When we need to refer to the spins on the components
of $\gamma$ they will be called $j_i$.
Note also that as we are working in ${\cal S}^3$ there are no
kinks.  

Now, let us assume that we have
an initial network $\rho=\{ Gamma , \gamma \} $ and
let us consider the corresponding space
${\cal F}_{\rho}$.  
The key point is that the links do not evolve, so that the
general solution can be constructed from the solutions
we have already described by decorating the 
initial trivalent component $\Gamma$
with non-intersecting loops.

Let us then consider 
a general element of ${\cal F}_{\rho}$.  It consists of states
where the network $\Gamma$ has been dressed, while
the link $\gamma$ remains unchanged.   As a result,
for every state $|\Gamma, n ,\alpha > \in {\cal F}_\Gamma$
and every way to extend $\Gamma$ to $\rho$ by adding
a link $\gamma$, 
there is a state 
$|\{ \Gamma, \gamma \} ,n ,\alpha > \in {\cal F}_{\{\Gamma ,\gamma \}}$.
The correspondence extends to the physical states as well, so
that for every solution in ${\cal P}_\Gamma$ given by 
\f
| \Gamma , Z> = \sum_n\sum_\alpha C^Z [n, \alpha ]
| \Gamma ,n ,\alpha>
\label{tetsol}
\ff
 labeled
by $Z$ and every way to add a link $\gamma$, there is a new solution
\f
|\{ \Gamma , \gamma \} , Z > = 
\sum_n\sum_\alpha C^Z [n, \alpha ]
| \{ \Gamma , \gamma \} ,n ,\alpha >   .
\label{linksol}
\ff
with the same coefficients $C^Z [n,\alpha ]$. 
The fact that the dressing procedure can be visualized 
as spreading each node out into a disk guarantees
that in each term in the sum \ref{linksol} the components
of $\gamma$ pass in between the disks, and not
among the new edges.  
As a result, the diffeomorphism classes
$\{ \Gamma , \gamma \}$ are relevent also for the whole
space of solutions ${\cal P}_\rho \subset {\cal F}_{\rho}$.
 
A final comment is that the surface $S$ 
may be self-intersecting.
This is allowed in general, as one may begin with a 
trivalent graph which is not planar, in which case its
decorations will describe self-intersecting surfaces.
However, the same reasoning we have just given shows
that becaues the dressing procedure  consists of the
spreading out of trivalent nodes into disks, there is always
left a hole through which other
elements of the network may pass.   Thus, there is no
obstruction to constructing the space of solutions
${\cal F}_\Gamma$ associated to a trivalent network whose
spanning surface is self-intersecting.

\section{The string state}

 Now that we have identified a set of collective coordinates 
that involve surfaces and degrees of freedom defined on them we
may see if we can use them to describe quantum states that might
be interpreted in terms of strings or membranes.  
 
To do this we may make use of a
particularly natural observable which is defined
on this space of states.  This is the area of a non-self-intersecting surface
$S$ embedded in a link $\gamma$ whose components 
$\gamma_i$ are colored by spins $j_i$.
It is defined by\cite{ls-review,volume1}
\f
{\cal A}[S] |\gamma , j_i> 
= l_{Planck}^2 
\sum_{i}  Int^+ [S, \gamma_i ]
\sqrt{j_i (j_i+1) }| \gamma ,  j_i >
\label{area}
\ff
We may note that this definition makes use of the unoriented
intersection number $Int^+ [S, \gamma_i ]$, of the surface
and the link, which is always positive,
and the $j$ in $\sqrt{j(j+1)}$ is the spin of the component of the
link at the intersection.

We would like to make use of this definition to define an
area associated with a generic physical quatum state 
$| \{ \Gamma , \gamma \} , Z >$.  This can be done in
the case that the initial spin network $\Gamma$  spans
a surface $S$ which is non-intersecting.  In this case
all the contributions to the area of $S$ come from the 
links $\gamma$, which are not part of the linear combinations
of states evolved from $\Gamma$ by the action of ${\cal H}(N)$,
out of which the solutions are built.  

To define the area we must take into account the fact that 
with the rules as defined here the spin networks that make
up the solutions do not span the whole surface $S$, but only the
disks we described in the previous section.  This means that there
is an ambiguity in how the linking among the spinnets  in
the diffeomorphism classes $\{\{ \Gamma , \gamma \} \}$
may be considered to intersect the surfaces $S$.  To resolve this
we will {\it define } the intersections of the surface $S$ with the
components of the link $\gamma$ in terms of the linking
of $\gamma$ and the initial network $\Gamma$.  To do this, let
us divide the surface $S$ spanned by $\Gamma$ into polygons,
$S_\alpha$ whose boundaries are the edges of $\Gamma$.
We then define the unoriented intersection number (which is
what is required for the area formula) as,
\f
Int^+ [S, \gamma ] \equiv \sum_\alpha \left |L[\partial S_\alpha, \gamma ]
\right |
\ff
where $L[\partial S_\alpha, \gamma ]$ is the standard 
Gauss linking number.  This definition tells us that  
intersections of the link $\gamma$ with the surface $S$ are
counted only if they are meaningful in terms of the diffeomorphism
classes of the states that make up the solutions\footnote{This definition
has the advantage that the subtelties associated with the presence
of nodes in surfaces\cite{AL1} are not relevent.}.

 Given this, we may define an operator $\hat{\cal A} $
which measures the area
of physical states $| \{ \Gamma , \gamma \} , Z >$.  We define
it such that
\f
\hat{\cal A} | \{ \Gamma , \gamma \} , Z > = 0
\ff
when the surface $S$ spanned by $\Gamma$ is self-intersecting
and, otherwise,
\f
\hat{\cal A} | \{ \Gamma , \gamma \}  , Z > = 
a [\{ \Gamma, \gamma \} ]
| \{ \Gamma , \gamma \} , Z > 
\ff
where, the area $a [\{ \Gamma, \gamma \} ]$ is defined by
\f
a [\{ \Gamma, \gamma \}] = 
\sum_\alpha \sum_i 
\left | L[\partial S_\alpha, \gamma_i ] \right | l_{Planck}^2
\sqrt{j_i (j_i+1) }  .
\ff

With these definitions, we may now define a special
physical state, $\Psi_{String} [\rho ]$ in the
spin network basis.  
Because of the general form of solutions we found in the last section,
any solution has the form
\f
\Psi [\rho ] = 
\Psi (\{ \Gamma ,\gamma \} , Z  ) .
\label{firstform}
\ff
We choose to define $\Psi_{String} [\rho  ]$
so that
\f
\Psi_{String} [\{ \Gamma ,\gamma \} , Z ]=0 
\label{restriction} 
\ff
whenever $  \Gamma $
describes a self-intersecting
surface and, otherwise, 
\f
\Psi_{String} (\{ \Gamma , \gamma \} , Z   )=
e^{{i g^2 } a [ \{ \Gamma , \gamma  \}  ] / l_{Planck}^2}  .
\label{stringstate}
\ff
Note that the state is parameterized by a dimensionless
coupling constant $g$, which introduces, as we will
see, the string tension
\f
\alpha^\prime = {l_{Planck}^2 \over g^2}
\label{stringtension}
\ff

\section{The classical limit and the emergence of a string theory}

We now consider what can be said about the classical
limit of the state \ref{stringstate} we have just defined.
To do this we will make use of the dependence of the
state on the non-intersecting link  $\gamma$.  We
may recall that there are special links $\gamma$ that
may be associated to classical slowly varying three
metrics, using the weave construction\cite{weave}.

It is easiest if we think of the state in the form \ref{firstform}
as a function on spin networks, and not just on their diffeomorphism
invariant classes.  (We are free to do this as the state exists in
both the kinematical state space and the space of diffeomorphism
invariant states.)   

We will use the definition of weaves given in
\cite{weave}  
in which the correspondence to a classical
spatial metric $q_{ab}$ is given in terms of the areas of large 
surfaces.   However, before going any further we
must discuss two issues concerning the 
interpretation of weave states.  

The first issue arises because 
the interpretation of non-intersecting weaves\cite{weave}
is problematic, because these states have the special property that
they are simultaneously in the kernel of the Hamiltonian
constraint and the volume operator\footnote{This is true for all
known forms of the volume operator, including the 
$q$-deformed operator.}.  
This implies
\f
\hat{\cal K} |\gamma  > = 
[{\cal H}(N) , \hat{\cal V} ]|\gamma > =0
\label{kisnot}
\ff
where ${\cal K}=\int d^3x \sqrt{q}K$, where $K$ is the
trace of the extrinsic curvature.  Still, stranger, one can
show  using Thiemann's methods\cite{qsdi} 
that\footnote{This is where the restriction to
states without kinks is used.}
\ref{kisnot} implies
\f
\hat{T}(x) |\gamma  > =0
\ff
where $T(x)$ is the kinetic term in the Minkowskian 
signature Hamiltonian constraint
which is quadratic in the extrinsic curvature $k_{ai}(x)$.  
This implies that
the state is static, i.e. that if it corresponds to a classical
spacetime it is one in which $k_{ai} (x)=0$.   

Thus, the background geometry defined by a weave
without intersections is one in which there is no
evolution so the  time dimension
plays no role.  It is then natural to interpret the classical
limit of such a state as describing a three dimensional
Riemannian geometry.  One can say that since
no degrees of freedom can depend on the time dimension, 
what we have here is the quantum analogue of a static
geometry.  As far as dynamics is concerned it is natural to
interpret this as saying that the quantum state describes
a world that has spontaneously compactified, or better
simply eliminated, the time direction,
so that it describes a world of three 
rather than four dimensions.
 
The second issue concerns the possible ambiguities in the
correspondence given in \cite{weave}
between a slowly varying three
metric $q_{ab}$ and a link $ \gamma $.
First,  if there is one link $\gamma $ that corresponds
to a given $q_{ab}$ there are many, so there
is no unique map ${\cal W}: q_{ab} \rightarrow \gamma_q$.

For the following we will find it useful to resolve this
ambiguity and fix a definition of a weave that
results in a unique  weave map
\f
{\cal W}: q_{ab} \rightarrow  \gamma  
\ff
To do this we need to give a definite construction
that results in a unique  weave $\gamma $ given any
slowly varying
three metric.  The prescription does not have to
be optimal, any construction that gives a weave
that satisfies the tests of agreements of areas of
surfaces given in \cite{weave} 
will do.  It is clear from the construction
in \cite{weave} that this may be easily done, for
example, the random selection of centers and angles
prescribed there may be specified according to
a particular random number generator.  In this
case there is a unique prescription given a
three metric that produces a weave, such that measurements
of areas of all surfaces of the two agree up to terms small
in Planck units.  

Given such a choice, we have, for all slowly 
varying\footnote{according to the definition 
of \cite{weave}} metrics
$q_{ab}$ a uniquely specified $\gamma [q_{ab}]$.

In essence, what we have done is identified a further set of 
collective coordinates associated with the dependence of the
state on the non-intersecting link $\gamma$ 
constain in a general trivalent
spin network $\rho$.  These collective coordinates are
a slowly varying three dimensional metric.  
We must emphasize that, unlike those considered so far, 
these collective coordinates 
are not generally valid, as the
weave correspondence is only sensible when the link has the
property that it is the image of some slowly varying metric under
the weave map.  But it is sufficient to extract the physics of a 
sector of the theory, which is within the range of validity of the weave 
correspondence.   The situation is perhaps analogous to
how the physics of phonons may be extracted by constructing an
effective field theory.
The atoms in a system must be condensed into a solid for 
a description in terms of photons to be meaningful, so phonon degrees
of freedom are collective coordinates only in a portion of the 
Hilbert space of the fundamental theory. Similarly,  a description
in terms of slowly varying geometry is only possible in a portion of the
state space of quantum gravity.

Now assume that we have a surface $S^\prime$
imedded in three dimensional space.  We would 
like to extend the weave map  so that it   associates
to the pair $(q_{ab} , S^\prime)$ a spin network which
consists of  an imbedding of a trivalent spinnet $\Gamma $ that
spans $S^\prime$ in the link $\gamma $ that 
represents $q_{ab}$.    This extended weave map will
satisfy, 
\f
{\cal W}:   (q_{ab}  , S^\prime ) 
\rightarrow ( \Gamma , \gamma  )_{q,S^\prime}
\ff
 where we have picked the imbedding of $\Gamma$ in
the link $\gamma  $ such that
\f
a [ \Gamma , \gamma    ] = {\cal A}^{class}[q_{ab}, S^\prime ]
+ O (  l^2_{Planck}) 
\ff
where ${\cal A}^{class}[q_{ab}, S^\prime ]$ is the classical formula
for the area of $S^\prime$ as a function of $q_{ab}$.
   
The weave map can then be further extended to give a map
from pairs $(q_{ab},S^\prime )$ and physical states,
$|\{ \Gamma \cup \gamma \} , Z, j_i >$ such that
\f
{\cal W} : (q_{ab},S^\prime ) \rightarrow 
| \{ (\Gamma , \gamma )_{q,S^\prime }  \} , Z^* >
\ff
where $Z^*$ is some arbitrary value of the other collective 
coordinates.

We can now use the weave map to define an effective quantum
state which is a function of the pair $(q_{ab}, S^\prime )$.  
This should be interpreted as any effective quantum theory:  it describes
a subset of the  degrees of freedom of the physical states which are
relevent for the long distance physics of a particular sector of the theory.
It is well defined only when the pair $(q_{ab}, S^\prime )$ is slowly
varying, in the sense defined in \cite{weave}, so that the weave map
is well defined.   

Given a physical state $\Psi [\{ \Gamma , \gamma \} ,j_i , Z ]$ the 
effective quantum state is defined by,
\f
\tilde{\Psi} (q_{ab}, S^\prime) \equiv \Psi \left [
 \{ (\Gamma , \gamma )_{q,S^\prime }  \} , Z^* \right ]
\ff
As it is defined only for metrics and surfaces that are slowly
varying on the Planck scale, the  functional 
$\tilde{\Psi}(q_{ab}, S^\prime )$  provides a way to
extract a semiclassical limit  from a general state $\Psi [\rho ]$.

The next question to ask is whether classical equations of motion might
govern the approximate dependence of the effective state
$\tilde{\Psi}(q_{ab}, S^\prime )$.   If this is the case we will have
defined a good classical limit for the 
physical state $\Psi [\{ \Gamma , \gamma \} , Z ]$.
It is easy to see that this is the case for the string state.  We have, for
slowly varying surfaces and three metrics,
\begin{eqnarray}
\tilde{\Psi}_{string} (q_{ab}, S^\prime) &\equiv &
\Psi_{string} \left [
( \{ (\Gamma , \gamma )_{q,S^\prime }  \} , Z^* \right ]
\nonumber \\
&= &e^{\imath {\cal A}^{class} [q_{ab} S^\prime ] /\alpha^\prime}
+ O (l_{Planck}^2 / {\cal A}^{class} [q_{ab} S^\prime ] 
\label{stringstatec}
\end{eqnarray}

This state may be interpreted as giving a string theory,
described by the motion of the surface $S^\prime $
in the background three dimensional Euclidean spacetime
described by $q_{ab}$.   To see this note that the
semiclassical limit is given by the
principle of stationary phase.  Applied to \ref{stringstatec}
we have, holding the background geometry fixed,
\f
{\delta  {\cal A}^{class} [q_{ab} , S^\prime ] \over 
\delta X^a (\sigma ,\tau ) } =0
\label{stringeqn}
\ff
where $X^a (\sigma , \tau )$ are the coordinates of the
embedding of the surface $S^\prime $.  
These are the equations of standard string theory in
a background $q_{ab}$ (with the other background
fields, the dilaton and antisymmetric tensor
field vanishing.)  

\section{Discussion}

There are a number of questions we may ask about
the interpretation of the result we have just found.   

The first is about the domain of validity of
the approximation that gives rise to the interpretation
of the state \ref{stringstate} as describing the propagation
of strings in a static background.  The weave interpretation
requires a scale $R >> l_{Planck}$ on which the metric
and all surfaces used in the construction are slowly 
varying\cite{weave}.  We may note that there is naturally
a second scale in the problem, which is 
$l_{string} = \sqrt{\alpha^\prime}$.  When 
$l_{string} >> l_{Planck}$ it may serve as $R$, in this case the
domain of validity of the stationary phase approximation
may agree with that of the weave construction.  This is also
the case in which the degrees of freedom of string theory are
cut off at a scale much larger than the Planck scale.  

A second question we may ask is which string theory we have
found.  This depends on the dynamics of the other
collective coordinates, which include the choice of
$\Gamma$, the original skeletonization
of $S$, and the $Z$'s.  It is important to stress  that
the physics of these additional degrees of freedom on the 
surfaces will depend greatly on the form of the Hamiltonian
constraint.  For example, with the rules we have so far
defined the $Z$'s  cannot be massless on the worldsheet, as
their correlations will be confined to regions of the worldsheet,
reflecting the problem of the  confinement of correlations
discussed in \cite{instability}.  However, with the definition
given in the appendix this problem is avoided, so that
these degrees of freedom may have long ranged
correlations on the surface.
In this case it would be very interesting to try to extract a conformal
field theory to describe these additional collective coordinates.

Alternatively, it is possible to extend the
construction by adding additional degrees of freedom to the
non-perturbative theory as well.   One case in which this
can be done rather easily is that of 
the antisymmetric tensor gauge field (or
Kalb-Ramond field).  As fields of this kind play an important role
in string theory, it is interesting to investigate them in this 
context, as is briefly done in the next section.   

\section{The Kalb-Ramond field and the surface representation}

As described previously in \cite{me-surfaces}, the 
Kalb-Ramond field turns out to fit rather neatly into
non-perturbative quantum gravity.  This is because there
is a surface  representation\cite{rodolfo-surfaces,me-surfaces}
for antisymmetric tensor gauge fields $B_{ab}$, in which
gauge invariant functionals are labeled by closed surfaces 
$S$ by
means of
\f
\tilde{\Psi}[S] = \int d\mu(B_{ab}) e^{k\int_{S} B}.
\label{transform}
\ff
 
As in the loop representation, any Hamiltonian formulation
involving gauge invariant functionals such as $H=dB$ may
be rewritten in this surface 
formalism\cite{rodolfo-surfaces,me-surfaces}.  Furthermore,
the term of the Hamiltonian constraint for $B_{ab}$ is
polynomial when written with density weight two, as is
suitable for coupling to the Ashtekar form of 
general relativity\cite{me-surfaces}.

When combined with the loop representation for quantum
gravity one has
a kinematical state space defined on diffeomorphism
invariant functionals of spin networks and surfaces,
whose states are of the form $\Psi [ \{ \Gamma , S \} ]$.
We see that by including $B_{ab}$ we have this form of the
state space at the kinematical level, without the necessity
to either restrict to the trivalent sector of solve the dynamics.
We may then posit directly the ``string state",
\f
\Psi_{string,B} [ \{ \Gamma , S \} ]\equiv 
e^{g^2 A[ \{ \Gamma , S \} ]/l_{Planck}^2}
\ff
It is interesting to note that taking into account \ref{transform}
the phase factor in the semiclassical limit  is
proportional to the classical string action,
$S_{string}= \int_\Sigma ( \sqrt{h} + B )$.
This has properties similar to \ref{stringstate}, in particular,
in the appropriate classical limit it describes extremal two
dimensional surfaces imbedded in a background metric
defined by the spin networks.  However, it should be pointed
out that $\Psi_{string,B}$ is not a solution of the Hamiltonian
constraint including the terms for $B_{ab}$ (described in
\cite{me-surfaces}).  Thus, one cannot extend the full interpretation
given to \ref{stringstate} to this case.

\section{Conclusions}

It is clear that the results described here are at most an indication
of a possible connection between non-perturbative quantum
gravity and string theory.  
To develop them further two kinds 
of implications may be explored.
The first is to the possibility of a non-perturbative formulation
of string theory.  
In this connection it is interesting to recall  a suggestion of Witten
that a four dimensional non-supersymmetric theory
might be the strong coupling limit of a three dimensional 
supersymmetric string theory\cite{34}.   This is based
on an analogy with the conjectured relationship betwen ten
dimensional string theory and eleven dimensional supergravity\cite{1011}.
While the nature of the four dimensional theory that follows from
the strong coupling limit of $3D$ string theory is unclear, if the
analogy holds there should be a field theory in four dimensions
that plays the same role of eleven dimensional supergravity
(That is it is at least the classical limit of the four dimensional
theory).
>From the non-perturbative point of view, this theory must be 
diffeomorphism invariant, as it will be a theory of gravity.
We also know from Witten's argument that this theory will have
zero cosmological constant, $\Lambda$ and no dilaton.

Moreover, if Witten's conjecture is correct than the relationship
might also work the other  way as well, in which case this
four dimensional field theory should have 
a limit which is described by a three
dimensional supersymmetric string theory.  

>From the non-perturbative point of view, we may try to represent
this theory in terms of spin networks, as they provide a
very general language for describing the non-perturbative
kinematics of any theory whose degrees of freedom may be 
defined in terms of a connection.
Indeed, the simplest possibility would be that the 
conjectured theory is just
quantum general relativity, perhaps coupled to some matter
fields, quantized non-perturbatively and 
tuned to $\Lambda =0$.
This might seem quite implausible, but we have
seen here that it is indeed the case that  
a sector of that theory has a semiclassical limit that describes a 
string theory in three dimensional 
spacetime\footnote{We may note that at the purely classical
level there is a degenerate sector of general relativity,
formulated in terms of Ashtekar variables, that describes
a three dimensional theory\cite{jljw} It is not known if there
is any relationship between this and the present results.}.  Further, 
the solutions we have been describing are
for $\Lambda=0$ and would not have existed for $\Lambda \neq 0$.
We may also note that the theory we have studied has no
dilaton and that the non-perturbative 
dimensional reduction mechanism that emerges here does not yield a 
compactification radius, both characteristics of the four dimensional
theory Witten conjectures as the strong coupling limit of the
three dimensional string theory.  
 
Now, numerical simulations\cite{AM,Ambjorn,reggemodels,review-triangles},
together with general renormalization group arguments suggest strongly
that quantum general relativity at bare $\Lambda=0$ cannot have a
continuum limit that can be described in terms of a four dimensional 
field theory.   This is consistent with what we have found so far,
which is that there is a sector of that theory which has a continuum
limit which may be described in terms of a three dimensional 
string theory.  Of course, 
this is still a long way from realizing Witten's picture,
according to which the three dimensional string theory whose strong
coupling limit gives a four dimensional field theory is supersymmetric.
So far there is no evidence that the string theory found here
is supersymmetric.  However, it is not impossible that 
a form of the Hamiltonian constraint exists 
such that the additional collective coordinates describe 
fermionic
conformal fields that, together with the embedding coordinates,
realize worldsheet supersymmetry. 

Whether or not this conjecture holds, it would be also interesting to see
if the mechanism discovered here extends to supergravity in
$11$ dimensions.  
To investigate this, we should first study canonical
quantization of supergravity in $10$ and $11$ dimensions.  
The required canonical formalisms do not, so far as
I know, so far
exist, but they may be developed.  This would also be interesting as
the study may reveal kinematical structures that could underlie
a non-pertubative string theory. 
Indeed,  one might conjecture that an extension of these results would
show that $\cal M$ theory was in fact nothing more than a
non-perturbative quantization of $11$ dimensional supergravity.
But even if this is not the case it is possible that the phenomena
described here might be found to 
apply to that case and illuminate the physics
of $\cal M$ theory.

The second set of implications of these results are to non-perturbative
quantum gravity itself.  This study shows that non-trivial structures
may emerge from the solution of some forms of the constraints,
which can be described in terms of 
collective degrees of freedom.  Such structures may
be essential for understanding crucial questions of physical
interest such as the inner product and the continuum limit.  For
example, it may be that two surfaces continue to play a key role
in the parameterization of solutions, even when we lift the restriction
to trivalent states.  One question of definite physical interest is
the behavior of the new degrees of freedom associated with
surfaces, and their dependence on the choice of regularization
procedures.

Still another direction to extend these results is from a canonical
to a path integral formulation.   The  
new path integral formulation
of Riesenberger\cite{mikenew}, and 
Riesenberger and Rovelli\cite{mikecarlo} suggests that 
fluctuations of surfaces may play a role in a four dimensional
covariant
perturbation theory built around semiclassical states.   

However, to build a useful bridge between string theory and 
non-perturbative quantum gravity the main obstacle to be
overcome is the
fact that each is built from structures that are simplest in a
particular dimension: ten in one case, four in another.   If there
is to be such a bridge it will be likely based on the existence
of structures associated with $10$ or $11$ dimensions that
play roles analogous to those of 
spin networks and self-duality in four.

\section*{ACKNOWLEDGEMENTS}

This paper grew out of work with Roumen Borissov and Carlo 
Rovelli on the Hamiltonian constraint, and I am grateful to them
for many discussions and suggestions.  
Correspondence with Thomas Thiemann about the form of
the Hamiltonian constraint were very useful in completing
this paper.  I am grateful also to Stuart Kauffman for
encouraging me to think about the dynamics of spin networks
in the langauge of dynamical systems.
This work was begun while I
was a guest of The Rockefeller University in New York and
SISSA in Trieste,  and I would like to thank Nick Khuri
Mauro Carfora for hospitality.  This work
has also been supported by the NSF grant  PHY-93-96246.

\section*{APPENDIX}

Here I discuss some details concerning the regularizations of the
Hamiltonian constraint that satisfy the rules used.

A form of the Hamiltonian constraint that satisfies the first
set of rules, {\bf R1-3} was described 
in \cite{ham1,roumen-ham,ham2}.  One must amend the process
described there in that the aim is to construct a Hamiltonian
constraint, which is not diffeomorphism invariant, rather than
a Hamiltonian, which must be.  To do this one 
proceeds as follows.  One defines the regulated operator acting
on spin-network states as (compare eq. 5 of \cite{ham1})
\begin{eqnarray}
\hat{\cal C}^{\delta \epsilon}(N) |\Gamma > &\equiv&
\int d^3 x N(x) \hat{\cal C}^{\delta \epsilon}(x) |\Gamma >
\nonumber \\
&\equiv&
\int d^3 x N(x) \int d^y \int d3 z f^\delta (x,y) f^\delta (x,z)
\hat{\theta}_{ij}^{-1}
\nonumber \\
&& \times \left  (
\hat{T}^{{a}{b}} \left [
\gamma_{x,\hat{a}\hat{b}, \epsilon^2 \hat{\theta}_{ij}}
\cdot \gamma_{xy}\cdot h(\hat{a})\gamma_{yx}
\cdot \gamma_{xz}\cdot h(\hat{b})\gamma_{zx}
\right ]  \right.    \nonumber \\
&&  - \left. \hat{T}^{{a}{b}} \left [
\gamma_{x,\hat{a}\hat{b}, \epsilon^2 \hat{\theta}_{ij}}^{-1}
\cdot \gamma_{xy}\cdot h(\hat{a})\gamma_{yx}
\cdot \gamma_{xz}\cdot h(\hat{b})\gamma_{zx} \right ]
 \right ) |\Gamma >
\label{reg}
\end{eqnarray}
Here the notation is that of  \cite{ls-review,ham1,roumen-ham}, 
and all smearing
functions $f^\delta (x,y)$
depend on an arbitrary flat background metric
$h^0_{ab}$.  
$\gamma_{x,\hat{a}\hat{b}, \epsilon^2 }$ is a loop based at $x$ in
the $\hat{a}\hat{b}$ plane with area $\epsilon^2$,
$\gamma_{xy}$ is a straight line (in the background metric)
from $x$ to $y$ and $ h(\hat{a})$ means that is where you
make the insertion of the hand with index $\hat{a}$.
The only non-standard thing is that, as described in
\cite{ham1,roumen-ham,ham2} there is an additional
operator dependence which measures the angle
$\theta_{ij}$ between the $i$'th and $j$'th tangent vectors
incident at $x$.    This explicit operator dependence is
necessary to cancel a factor of $\theta_{ij}$ that arises
in the integrals over regulators.  The use of such explicit
additional operator dependence in the regulator is a
cleaner way to describe what is sometimes called
``state dependence of the regulator" and is described in
more detail in \cite{ham2}.

After some steps that parallel those 
in \cite{ham1,roumen-ham,ham2} we arrive at
\f
\hat{\cal C}(N)|\Gamma > = \sum_n \sum_{ij}{N(n) \over \delta 
\epsilon^2}
{3 \over 10 \pi} (16 \pi l_{Planck}^2 )^2 ij 
|\Gamma ** \gamma_{x,ij,\epsilon^2 \theta_{ij}}>
\ff
where the sums are over the nodes of $\Gamma$ labeled by $n$
and pairs of non-collinear edges $i$ and $j$ at $n$.
We then define the {\it renormalized} operator by
\f
\hat{\cal C}^{ren}(N)|\Gamma > \equiv \lim_{\epsilon \rightarrow 0}
\lim_{\delta \rightarrow 0} \epsilon^2 \delta 
{\cal C}^{\delta \epsilon}(N) |\Gamma >
\ff
Here, as described in more detail in \cite{ham1,roumen-ham,ham2}
the first limit is taken in the kinematical state space while the
second is taken in the space of diffeomorphism invariant states,
with the loop chosen to lie in a triangle bordered by the 
edges $i$ and $j$ (more additional operator dependence built
from $\dot{\gamma}(x)$.)

The result is an operator satisfying the rules 
{\bf R1-R3}.   

The next step is to modify this operator to eliminate the action
on kinks, so that it satisfies rule {\bf R3'} instead of {\bf R3}.
As we learn from \cite{qsdi,qsdii} one way to change the
class of nodes an operator acts on is to change its density weight.
Given that we do a renormalization at the end, and so end
up in any case with a non-diffeomorphism invariant operator,
there is no objection to doing this.  
We then modify the Hamiltonian constraint to,
\f
{\cal C}^\prime (N) \equiv \int d^3x N^\prime (x) {\cal C}(x)
{q(x)}
\ff
Where $N^\prime (x)$ is now a density of weight minus four.
In order that the density $q$ not annihilate the trivalent
vertices we are also going to have to $q$-deform the formalism,
as in \cite{sethlee,rsl}.  Before discussing this we attend to 
the details of the regularization.
We define the corresponding operator by keeping this
order and regulating each piece separately
\f
\hat{\cal C}^\prime_{\delta \epsilon L} (N) |\Gamma >
\equiv \int d^3x N^\prime (x) \hat{\cal C}^{\delta \epsilon}(x)
{\hat{q}(x)}^L  |\Gamma >
\ff
where $\hat{\cal C}^{\delta \epsilon}(x)$ is defined in \ref{reg}
and the regulated density $q(x)^L$ is defined as in 
\cite{volume1,rsl}.  We take a cube of linear size $L$ in the
background metric centered at $x$ and define
\f
\hat{q}^L\equiv {1 \over 2^7 3!L^6 }\hat{W}^{x,L}
\ff
where $\hat{W}^{x,L}$ is the operator defined by eq. 10 of
\cite{volume1} for a the cube of size $L$ around $x$.
We will then define the renormalized operator to be
 \f
\hat{\cal C}^{\prime ren}(N)|\Gamma > \equiv 
\lim_{\epsilon \rightarrow 0}
\lim_{\delta \rightarrow 0} 
\lim_{L\rightarrow 0} \epsilon^2 \delta L^6
{\cal C}^{\delta \epsilon}(N) |\Gamma >
\ff
with the limits taken in the order indicated.

We then work in the $q$-deformed formalism\cite{sethlee,rsl} 
based on $q$-deformed spin networks\cite{lou-sn}.  In that
formalism all trivalent nodes are eigenstates of the operators
$\hat{W}$ defined on every box small enough that it encloses
only one of them.  Let us call the eigenvalue $w(i,j,k)$, where
$i,j$ and $k$ are the three
spins incident on it; it is computed
in \cite{rsl}.  These eigenvalues are non-vanishing, without regard
for any linear dependences among the tangent vectors of the
incident edges.  In this respect (as well as in the $q$-deformation)
it differs from the operator defined in \cite{AL1}.
At the same time $\hat{W}$ for any box
annihilates kinks.  The action of the modified renormalized
operator is then to ignore the kinks and, when acting
on trivalent nodes, multiply the coefficients 
$A^{\pm \pm^\prime }(j,k)$  by $w(i,j,k)$.

The result is an operator which satisfies the modified
rule {\bf R3'} instead of {\bf R3}.

There is one more form of the Hamiltonian constraint that gives
that I want to discuss.  This is a
regularization defined directly in the diffeomorphism
invariant spin-network language.  As discussed in more
detail in \cite{instability} the idea is that the form
of the regulated operator that represents the Hamiltonian
constraint need not be derived from a regularization
procedure.  It is sufficient that it agree, when applied to
a non-diffeomorphism invariant state in the connection
representation and evaluated for connections that are
slowly varying (in the topology of the graph) with a
form of the constraint that is derived from a point-split
regularization.  Such a prescription can be given directly
by a set of rules.  One such operator is given in \cite{instability}.
I give here an alternate form that has the property that
its action closes on the trivalent spin 
networks\footnote{For those familar with \cite{instability}
the only modifications are in the restriction to action
on trivalent nodes and  in the step where new edges are
created.}.  It is called ${\cal C}^\prime_{new}$.

\begin{itemize}

\item{\bf N1} $ {\cal C}^\prime_{new} (N)$ acts on 
an element $\Gamma$ of the
spin network basis at each pair of non-collinear
edges $e_1$ and $e_2$ of every trivalent 
node $v$ in the following way.
It finds the first nodes adjacent to $v$ along $e_1$ and $e_2$,
which will be called $v_1$ and $v_2$.  

\item{\bf N2} Suppose there is an edge joining $v_1$ and 
$v_2$, which will
be called $e_{12}$.  The action of $ {\cal C}^\prime_{new}$
 produces a sum of six terms in
which the colors along $e_1$, $e_2$ and $e_{12}$ which
we call $i,j$ and $k$ respectively are updated by $\pm 1$.
Each is multiplied by an amplitude 
$A_{\pm , \pm^\prime , \pm^{\prime \prime}}(i,j,k;r,s,t)$
which I give below.
Here we assume that each of the nodes is written in the form
in which the two edges in the problem are joined to a third
edge at a trivalent vertex with an edge with definite color.
The colors
associated with these edges for $v, v_1$ and $v_2$,
respectively, are $r,s$ and $t$.  
$\pm , \pm^\prime $ and $ \pm^{\prime \prime}$ refer
respectively to the updating of $i,j$ and $k$.  The
amplitude is then,
\begin{eqnarray}
A_{\pm , \pm^\prime , \pm^{\prime \prime}}(i,j,k;r,s,t) &=&
\pm^{\prime \prime} ij  
\left \{ {i i i\pm1}; {112}  \right \}
\left \{   {j j j\pm^\prime 1}; {112}  \right \}
\left \{   {i\pm 1 i r }; {j\pm^\prime 1  j 1 }  \right \}
\nonumber \\
&&\times 
 \left \{   {i\pm 1 i s }; {k\pm^{\prime \prime} 1  k 1 }  \right \}
\left \{ {j\pm^\prime 1 j t }; {k k\pm^{\prime  \prime}1 1 }\right \}
\nonumber \\
&& \times
{ \Theta(i,j,r) \Theta (j,k,t ) \Theta (i, k,s)  \over         
[r+1][s+1][t+1]   }
\end{eqnarray}

Here $\left \{ {i i i\pm1}; {112}  \right \}$ are the $6-j$ symbols,
and $\Theta(i,j,r) $ is the theta function defined in 
\cite{lou-sn,rsl,roumen-ham}.  The formula is written in
a way that is good for either the ordinary or $q$-deformed
case, so $[n]$ is the quantum integer \cite{lou-sn}, which is
equal to $n$ in the ordinary case.  

There is also the case in which there is in $\Gamma$ no
edge joining $v_1$ and $v_2$.  In this case the
operator adds a new edge with color $1$.  Here the definition
must differ from that given in \cite{instability} so that
no vertices are produced with more than $3$ incident
edges.  To do this we break theedge $e_1$   
joining $v$ to $v_1$ and  
and insert a new node $v^\prime_1$.  We do the same
thing to $e_2$ creating a new node
$v^\prime_2$ between $v$ and $v_2$.  
The two halves of $e_1$ and $e_2$
are each colored by the same spins that colored the edges
originally.  Then we join these two new vertices
with a new edge with color $1$, which we call $e_{12}$.
The topology of the edge is chosen so the loop
it forms with the segments of $e_1$ and $e_2$ links or intersects
no other edge of the network.
One then applies
the above formula  
with $r=i$, $s=j$, $k=0$ and
$\pm^{\prime\prime}=+$, producing in this case
four terms.  

What this combinatorial formula corresponds to is adding a
loop as usual to represent the $F_{ab}$ in the plane of the
tangent vectors of the two edges.  The combinatorics are as in 
\cite{ham1,roumen-ham,ham2} except that the new loop
is taken to go around the triangle $e_1,e_2,e_{12}$.  

\item{\bf N3}  To complete the definition of the operator
we must divide by the area of the triangle $e_1,e_2,e_{12}$.
We may note that, as determined by the area operator, this will
often vanish, but it may instead be defined using 
Thiemann's length operator\cite{tt-length}  as follows.
If we call $\hat{L}_1,\hat{L}_2$ and $\hat{L}_{3}$ 
the length operators of
the edges of a triangle $\Delta$ of a spin-network, we 
may define an operator that measures its area as,
\f
\hat{A}_\Delta^2  = {1 \over 4}\left ( 
{ \hat{L}^2_1\hat{L}^2_2 \over 2}+{ \hat{L}^2_1\hat{L}^2_3 \over 2}
+{ \hat{L}^2_2\hat{L}^2_3 \over 2}
-{\hat{L}^4_1 \over 4}-{\hat{L}^4_2 \over 4}-{\hat{L}^4_3 \over 4}
\right )
\ff
 where we have used the standard formula from Euclidean geometry
for the area.  (If the operators fail to commute we take symmetric
ordering.)\footnote{Note also that 
the formula for $\hat{L}$ of an
edge given by Thiemann in \cite{tt-length} is ultimately
expressed in terms of $6-j$ symbols, which means that
the $q$ deformation can be immediately written down
by using the $q$ deformed $6-j$ symbols defined in
\cite{lou-sn}.}That this will often yield a different answer than a 
direct measurement of the area is an inevitable consequence 
that we are working with a quantum field theory, in which
functional relationships between classical observables may not
be preserved.

We may then define this step as follows:  If there is a term
with no triangle corresponding to the three original edges we
do nothing.  If there is we multiply the state gotten by the first
two steps by the operator $\hat{A}^{-1}_{e_1,e_2,e_{12}}$.  
We need to define 
the
inverse so it is well defined in the case that the area is zero.
We do so by $\hat{A}^{-1} \equiv \hat{A}^{-2} \hat{A}$,
where $\hat{A}^{-2}$ is defined on the space orthogonal
to the kernel of $\hat{A}$,
so that terms that might contribute zero area are projected out.
 
\end{itemize}

As discussed in more detail in \cite{instability} this operator
may resolve the problem of bounded correlations.  The solutions
it generates will still be characterized by two dimensional
surfaces, but there will in general be no single graph in the
superposition of states which plays a special role as the
ancestor of the other graphs.  Instead, it is possible that this
rule will lead to a flow that is free on the space of all
spin networks $\Gamma$ that span a given two dimensional surface, $S$.
If true this will mean that the parameters $Z$ that distinguish
the different solutions on $S$ 
will not be restricted to particular regions, but
will depend on how the spin networks propagate over the
whole surface.    If so they must be be described
in terms of conformal fields on each $S$. Investigation of this
form of the theory is in progress.


\begin{thebibliography}{99}

 \bibitem{strings}M. Green, J. Schwartz and E. Witten {\it
Superstring theory} Cambridge University Press.

\bibitem{strings-bh}See, for example, A. Sen hep-th/9504147;
F. Larsen and F. Wilczek, hep-th/9511064;
A. Strominger and C. Vafa, hep-th/9601029;
C. Callen and J. Maldacena hep-th/9602043; 
A. W. Peet, hep-th/9506200;
J.Maldacena, {\it Black holes i string theory} Ph.D. thesis,
and references contained theirin.

\bibitem{stringduality}See, for example, A. Sen,
Int. J. of Mod. Phys. A9 (1994) 3707, hep-th/9402002; 
C. M. Hull and P. K. Townsend, {\it Unity of superstring
dualities}  Nucl. Phys. B438 (1995) 109,
 hep-th/9410167;
M. J. Duff, R.R. Kuri and J. X. Lu hep-th/9412184;
J. Harvey and A. Strominger hep-th/9504047.
 
\bibitem{1011}E. Witten,  {\it String theory dynamics
in various dimensions}  hep-th/9503124, Nucl. Phys. B443 (1995) 85-126. 

 
 \bibitem{abhay1}A.A. Ashtekar, Phys. Rev. Lett. 57 (1986)
2244; Phys. Rev. D36 (1987) 1587

\bibitem{fernando-real}J. Fernando Barbero, Phys. Rev.
D49,  6935 (1994).

\bibitem{renata-real}R. Loll, ``A real alternative to quantum
gravity in loop space" gr-qc/9602041.

\bibitem{tedlee}
T. Jacobson and L. Smolin, Nucl. Phys. B 299 (1988).

\bibitem{pl}P. Renteln and L. Smolin, Class. Quant.
Grav. 6 (1989) 275.

\bibitem{lp1}C Rovelli L Smolin: Phys Rev Lett 61 (1988) 1155; Nucl 
Phys B133, 80 (1990).

\bibitem{lp2}R Gambini A Trias: Phys Rev D23 ,  553
(1981); Lett al Nuovo 
Cimento 38,  497 (1983); Phys Rev Lett 53,  2359 (1984); Nucl Phys 
B278,  436 (1986); R Gambini L Leal A Trias: Phys Rev D39 ,  3127
(1989);  R Gambini: Phys Lett B 255, 180 (1991)   

\bibitem{carlo-review}C Rovelli: Class Quant Grav 8,  1613 (1991)

\bibitem{ls-review}L  Smolin: in {\it Quantum Gravity and 
Cosmology}, eds  J  P\'erez-Mercader {\it et al}, World Scientific, 
Singapore 1992  

\bibitem{aa-review}A  Ashtekar:  {\it Non perturbative canonical 
gravity}, World scientific, Singapore 1991.

\bibitem{BGP}B. Bruegmann, R. Gambini and J. Pullin,
Phys. Rev. Lett.  68,  431 (1992); Gen. Rel. and Grav. 251 (1993).

\bibitem{GP}R. Gambini and J. Pullin, 
{\it  Loops, knots, gauge theories and quantum gravity}
Cambridge University Press, 1996.

\bibitem{g5}A Ashtekar J Lewandowski D Marlof J 
Mour\~{a}u T Thiemann:  ``Quantization of diffeomorphism
invariant theories of connections with local degrees of
freedom", gr-qc/9504018, JMP 36 (1995) 519.

 \bibitem{weave}A. Ashtekar, C. Rovelli and 
L. Smolin, Phys. Rev. Lett. 69 ,  237 (1992).

\bibitem{volume1}C. Rovelli and L. Smolin
{\it Discreteness of area and volume in quantum gravity}
 Nuclear Physics B 442 (1995) 593.  Erratum: Nucl. Phys.
B 456 (1995) 734.

\bibitem{AL1}A. Ashtekar and J. Lewandowski, "Quantum
Geometry I: area operator" gr-qc/9602046.

\bibitem{L1}J. Lewandowski, "Volume and quantization"
gr-qc/9602035.

\bibitem{tt-length}T. Thiemann, ``The length operator in canonical
quantum gravity" Harvard preprint 1996, gr-qc/9606092.


\bibitem{AI}A. Ashtekar and C. J. Isham, 
 Class and Quant  Grav 9 (1992) 1069 

\bibitem{instability}L. Smolin, {\it  The classical
limit and the form of the hamiltonian constraint in non-pertubative
quantum gravity} CGPG preprint, gr-qc/9609??.

\bibitem{ks}I. Klebanov and L. Susskind, Nucl. Phys. B309
(1988) 175.





\bibitem{34}E. Witten, {\it Strong coupling and the
cosmological constant} hep-th/9506101.

\bibitem{ham1}C Rovelli L Smolin: Phys Rev Lett 72 (1994) 
446 

\bibitem{roumen-ham}R. Borissov, {\it Graphical evolution
of spin network states}, gr-qc/9606013.

\bibitem{ham2}R. Borissov, C. Rovelli and L. Smolin,
{Evolution of spin networks} preprint in preparation.

 
\bibitem{sn1}C. Rovelli and L. Smolin,  
``Spin networks and quantum gravity"  
gr-qc/9505006, Physical Review  

\bibitem{sn-roger}R Penrose: in {\it Quantum theory and 
beyond}  ed T Bastin, Cambridge U Press 1971;
in {\it Advances in Twistor Theory}, ed. L. P. Hughston and R. S. 
Ward,
(Pitman,1979) p. 301; in {\it Combinatorial Mathematics and
its Application} (ed. D. J. A. Welsh) (Academic Press,1971).

\bibitem{renata2}R. Loll, Nucl. Phys. B444 (1995) 619.

\bibitem{GP2}H. Fort,  J. Griego, R. Gambini, J. Pullin, 
{\it Lattice quantum gravity in
the loop representation}, in preparation.

\bibitem{qsdi}T. Thiemann, Quantum spin dynamics I,
Harvard preprint (1996), gr-qc/9606089.

\bibitem{qsdii}T. Thiemann, Quantum spin dynamics II,
Harvard preprint (1996), gr-qc/9606090

 \bibitem{sethlee}S. Major and L. Smolin, {\it Quantum deformation
of quantum gravity} gr-qc/9512020.

\bibitem{rsl}R. Borissov, S. Major and L. Smolin, {\it The
geometry of quantum spin networks} gr-qc/9512043.

\bibitem{dpr}R. DePietri and C. Rovelli, ``Geometry eigenvalues and
scalar product from recoupling theory in loop quantum gravity",
gr-qc/9602023.

\bibitem{me-surfaces}L. Smolin, Phys. Rev. D 49 (1994) 4028.

\bibitem{rodolfo-surfaces}R. Gambini,  
Phys. Lett. B 171 (1986) 251; P. J. Arias, C. Di Bartolo, X. Fustero,
R. Gambini and A. Trias, Int. J. 
Mod. Phys. A 7 (1991) 737.

\bibitem{mikenew}M. Reisenberger, {\it A left handed
simplicial action for euclidean general relativity}  gr-qc/9609002.

\bibitem{mikecarlo}M. Reisenberger and C. Rovelli, preprint in
preparation.

\bibitem{lou-sn}L. Kauffman and S. Lins {\it Tempereley-Lieb
Recoupling Theory and Invariants of 3-Manifolds}
Princeton University Press, 1994, and references therein.
 
\bibitem{kodama}H. Kodama, Phys. Rev. D 42 (1990) 2548.

\bibitem{boundary}L. Smolin, JMP Nov. 1995 gr-qc/9505028.

\bibitem{witten-cs}E. Witten , Commun. Math. Phys.
121 (1989) 351.

\bibitem{jljw}J. Lewandowski and J. Wisioewski, 
{\it 2+1 sector of 3+1 gravity} gr-qc/9609019.

\bibitem{AM}M.E. Agishtein and A. A. Migdal, Nucl. Phys. B385 
(1982) 395.

\bibitem{Ambjorn}J. Ambjorn, J. Jerkiewicz and C. F. Kristjansen,
Nucl. Phys. B393 (1993) 601; Phys.  Lett. B305 (1993) 208;
J. Ambjorn, Z. Burda, J. Jerkiewicz and C. F. Kristjansen,
Phys. Rev. d48 (1993) 3695.

\bibitem{reggemodels}H. W. Hamber, Nucl Phys. B (Proc. Supp.)
20 (1991) 728; 25A (1992) 150; B400 (1993) 347;
Phys. Rev. D45 (1992) 507; H. W. Hamber and R. M. williams,
Nucl. Phys. B415 (1994) 463.

\bibitem{review-triangles}For a review, see 
J. Ambjorn, J. Jerkiewicz and
Y. Watabiki, ``Dynamical triangulations, a gateway to quantum
gravity" NBI-HE-95-08, to appear in J. Math. Phys. Nov. 1995.

\bibitem{chopinlee}L. Smolin and C. Soo,  
{\it The Chern-Simons invariant as the natural time variable for
classical and quantum gravity}  
Nucl. Phys. B 327 (1995) 205.

\end{thebibliography}
\end{document}